\documentclass[12pt]{article}

\usepackage{amsmath}
\usepackage{amssymb}
\usepackage{graphicx}
\usepackage[cp1250]{inputenc}

\renewcommand{\[}{\begin{equation}}                               %
\renewcommand{\]}{\end{equation}}

\newcommand{\pd}{\partial}
\newcommand{\im}{\mathrm{Im\,}}
\newcommand{\R}{\mathbb{R}}
\newcommand{\C}{\mathbb{C}}
\renewcommand{\AA}{\mathcal{A}}
\newcommand{\ba}{\mathbf{a}}
\newcommand{\bA}{\mathbf{A}}
\newcommand{\BB}{\mathcal{B}}

\newcommand{\rD}{\mathrm{D}}
\newcommand{\re}{\mathrm{e}}
\newcommand{\HH}{\mathcal{H}}
\newcommand{\rmi}{\mathrm{i}}
\newcommand{\rJ}{\mathrm{J}}
\newcommand{\bk}{\mathbf{k}}
\newcommand{\bK}{\mathbf{K}}
\newcommand{\bp}{\mathbf{p}}
\newcommand{\rR}{\mathrm{R}}
\newcommand{\RR}{\mathcal{R}}
\newcommand{\bx}{\mathbf{x}}

\newcommand{\vk}{\varkappa}

\begin{document}

\title{A ``hybrid plane'' with spin-orbit interaction}

\author{Pavel Exner$^{1,2}$ and Petr \v{S}eba$^{2,3}$}
\date{\small 1) Nuclear Physics Institute, Czech Academy of Sciences,
25068 \v{R}e\v{z} near Prague, \\
2) Doppler Institute, Czech Technical University, B\v{r}ehov\'{a}
7, 11519 Prague, \\
3) University of Hradec Kr\'alov\'{e}, V\'{\i}ta Nejedl\'eho 573,
50002 Hradec Kr\'alov\'{e}, Czechia \\
\emph{e-mail: exner@ujf.cas.cz, seba@fzu.cz}\\ [3em] \textsl{In
memoriam Vladimir A. Geyler (1943-2007)}}

\maketitle


{\small \noindent In this paper we attempt to reconstruct one of
the last projects of Volodya Geyler which remained unfinished. We
study motion of a quantum particle in the plane to which a
halfline lead is attached assuming that the particle has spin
$\frac12$ and the plane component of the Hamiltonian contains a
spin-orbit interaction of either Rashba or Dresselhaus type. We
construct the class of admissible Hamiltonians and derive an
explicit expression for the Green function applying it to the
scattering in such a system.}

\setcounter{equation}{0}
\section{Introduction}

There is no doubt that in the person of Volodya Geyler, science
lost a bright personality, somebody who with safety and elegance
treaded forward at the boundary between mathematics and physics,
mastering the subtleties of the former and understanding deeply
the meaning of the latter. For the authors of the present paper
the sad news have a personal touch because the last talk he
announced bore the title \emph{Exner-\v{S}eba hybrid plane with
the Rashba Hamiltonian}; he passed away at the opening of the
conference in the Isaac Newton Institute in Cambridge where it had
to be presented.

We can only guess what Volodya intended to report but we decided
that the best way to honour his memory is to reconstruct what we
think could be the contents of this lecture. We take the ``hybrid
plane'' which we introduced twenty years ago and look what happens
if the particle living in it has a spin and is subject to
spin-orbit interaction in the plane part of the configuration
space; we consider the two interaction forms which were objects of
interest recently. We construct the class of admissible
Hamiltonians described by the boundary condition in the coupling
point between the plane and the halfline lead attached to it.
After that we derive an explicit expression for the respective
Green's functions using standard Krein's function technique, and
show how one can use them to derive properties of such system.

Since this is not intended to be an in-depth study, we will speak
mostly about the simple situation when there is no external field.
It is easy to extend the results to the situation when the
particle is subject to a homogeneous magnetic field in the plane.
We will comment on this case briefly hoping that this and other
ideas related to the problem will find a continuation.

\setcounter{equation}{0}
\section{Spin-orbit interaction}

Let us first recall how one describes a two-dimensional particle
with spin-orbit interaction; as announced we will pay most
attention to the simple case where there is no external field,
adopting the notation from \cite{BGP07}. For a particle with two
spin states the state Hilbert space is $\HH_\mathrm{plane}=
L^2(\R^2,\C^2)$ and its free motion is described by $\hat H_0=
\frac{1}{2m^*} \bp^2 \sigma_0$, where $p_j= -i\hbar \pd_j,\,
j=1,2,\,$ as usual, and $\sigma_0$ is the $2\times2$ unit matrix.
The spin-orbit interaction is introduced in different way: one is
the so-called \emph{Rashba Hamiltonian}
 \[ \label{RHam}
\hat H_\rR := \hat H_0 + \frac{\alpha_\rR}{\hbar} \hat U_\rR\,,
\qquad \hat U_\rR := \sigma_1 p_2 -\sigma_2 p_1 \,,
 \]
where $\alpha_\rR\in\R$ is the Rashba constant and $\sigma_j$ are
the usual Pauli matrices, the other is the \emph{Dresselhaus
Hamiltonian}
 \[ \label{DHam}
\hat H_\rD := \hat H_0 + \frac{\alpha_\rD}{\hbar} \hat U_\rD\,,
\qquad \hat U_\rD := \sigma_2 p_2 -\sigma_1 p_1 \,,
 \]
in which interaction strength is given by Dresselhaus constant
$\alpha_\rD\in\R$.

Since the choice of the units will not be important in the
following we get rid of the constants in the usual way introducing
$\bk:= \hbar^{-1} \bp$ and $\vk_j:= \hbar^{-2}m^* \alpha_J,\,
j=R,D$. Up to the multiplicative factor, $\hat H_\rJ =
\frac{\hbar^2}{2m^*} H_\rJ,\: \rJ=\rR,\rD,\,$ the both versions of
the Hamiltonian acquire then the simple form
 \[ \label{SOHam}
H_\rJ = H_0 + 2\vk_\rJ U_\rJ\,, \qquad U_\rR := \sigma_1 k_2
-\sigma_2 k_1\,, \quad U_\rD := \sigma_2 k_2 -\sigma_1 k_1 \,,
 \]
with $H_0:= \bp^2 \sigma_0$, which we shall use in the following.

As usual the properties of such a Hamiltonian are encoded in its
resolvent. The latter is known explicitly from the paper
\cite{BGP07}. By a nice algebraic trick, so characteristic for the
work of Volodya Geyler, the derivation is reformulated as a scalar
problem which involves the well know resolvent kernel $G_0(\bx,
\bx';z)= \frac{1}{2\pi} K_0(\sqrt{-z} |\bx-\bx'|)$ of the
Laplacian in $L^2(\R^2)$, where $K_0$ is the zero-order MacDonald
function. It leads to the expression
 \[ \label{SOGreen}
G_{\rJ}(\bx,\bx';z)=\left(\begin{array}{cc}
G^{11}_{\rJ}(\bx,\bx';z) & G^{12}_{\rJ}(\bx,\bx';z)\\
G^{21}_{\rJ}(\bx,\bx';z) & G^{22}_{\rJ}(\bx,\bx';z)
\end{array}\right)
 \]
with the diagonal elements
 \begin{eqnarray*}
 \lefteqn{G^{11}_{\rJ}(\bx,\bx';z)=G^{22}_{\rJ}(\bx,\bx';z)=
 \displaystyle\frac{1}{4\pi}\,\bigg[
 -\displaystyle\frac{\vk_{\rJ}}{\rmi\sqrt{-(z+\vk_{\rJ}^2)}}}
 \\ && \times
 \big( K_0(\zeta^+_{\rJ}|\bx-\bx'|)-K_0(\zeta^-_{\rJ}|\bx-\bx'|)
 \big) +K_0(\zeta^+_{\rJ}|\bx-\bx'|)+K_0(\zeta^-_{\rJ}|\bx-\bx'|)
 \bigg]
 \end{eqnarray*}
for both the $\rJ=\rR,\rD$, while the off-diagonal ones are
 \begin{eqnarray*}
G^{12}_{\rR}(\bx,\bx';z) &\!=\!&
\displaystyle\frac{\rmi(x_2-x'_2)-(x_1-x'_1)}{4\pi\,
\rmi\sqrt{-(z+\vk_{\rR}^2)}\,|\bx-\bx'|}\: \sum_{\nu=\pm} \nu\,
\zeta^\nu_{\rR}K_1\big(\zeta^\nu_{\rR}|\bx-\bx'|\big)\,, \\
G^{12}_{\rD}(\bx,\bx';z) &\!=\!& \displaystyle\frac{(x_2-x'_2)
-\rmi(x_1-x'_1)}{4\pi\, \rmi \sqrt{-(z+\vk_{\rD}^2)}\,|\bx-\bx'|}
\: \sum_{\nu=\pm} \nu\, \zeta^\nu_{\rD}
K_1\big(\zeta^\nu_{\rD}|\bx-\bx'|\big)\,,
 \end{eqnarray*}
and $G^{21}_{\rJ}(\bx,\bx';z)=\overline{G^{12}_{\rJ}(\bx',\bx;\bar
z\,)}$; the effective momenta are defined at that as
 \[ \label{effm}
\zeta^\pm_{\rJ}:=\sqrt{-(z+\vk_{\rJ}^2)}\pm \rmi\vk_{\rJ}
 \]

For the hybrid plane model which we will describe in the next
section we will need also the renormalized Green's function, i.e.
the diagonal value obtained after the subtraction of the divergent
term,
 \[ \label{Gren}
G^{\mathrm{ren}}_{\rJ}(z):=\lim_{\bx'\to\bx} \Big[
G_\rJ(\bx,\bx';z) +
\displaystyle\frac{1}{2\pi}\ln|\bx-\bx'|\sigma_0 \Big]\,;
 \]
notice that the limit is independent of the position $\bx$ in view
of the translational invariance of the Hamiltonian $H_\rJ$. By a
straightforward computation \cite{BGP07} one finds that the
off-diagonal elements vanish in the limit while
 $$
G^{\mathrm{ren}; jj}_\rJ(z)= -\frac{\vk_\rJ}{2 \rmi
\sqrt{-(z+\vk_{\rJ}^2)}}\big(
Q(\zeta^+)-Q(\zeta^-)\big)+\frac{1}{2}\big(
Q(\zeta^+)+Q(\zeta^-)\big)
 $$
with $Q(z):= \frac{1}{2\pi}\big(\psi(1)-\frac12\ln (-z)+\ln
2\big)$ expressed through the digamma function, hence the
renormalized Green's function can be written as
 \[ \label{Gren2}
 G^{\mathrm{ren}}_\rJ(z)= \frac{1}{2\pi}\bigg[
\psi(1)-\frac{1}{2}\ln\big(-\frac{z}{4}\big) +\frac{\vk_\rJ}{2\rmi
 \sqrt{-(z+\vk_{\rJ}^2)}}\,\ln
 \frac{\sqrt{-(z+\vk_{\rJ}^2)}
+i\vk_\rJ}{\sqrt{-(z+\vk_{\rJ}^2)}-\rmi\vk_\rJ} \bigg]\sigma_0\,;
 \]
recall that $-\psi(1)\approx 0.577$ in the above formula is
Euler's number.

The case when a homogeneous magnetic field $B=\frac{\hbar c}{e} b$
perpendicular to the plane is applied is treated in a similar way.
The momentum $\bk$ in the Hamiltonian (\ref{SOHam}) has to be
replaced with $\bK=\bk-\ba$ where $\bA= \frac{\hbar c}{e}\ba$ is
the vector potential associated with the field, and the Zeeman
term $\gamma b\sigma_3$ with $\gamma:= \frac12 g_*
\frac{m_*}{m_e}$ has to be added. The the reduction to the scalar
case works again and yields explicit expression for Green's
functions \cite{BGP07} in terms of confluent hypergeometric
instead of Bessel functions.

After this preliminary, we are ready to turn to our proper
subject.

\setcounter{equation}{0}
\section{Motion in the hybrid plane}

Let us now consider that the system has a mixed dimensionality and
its configuration space consists, as in \cite{ES87}, of a plane
described above to which a halfline lead is attached;
conventionally we place the junction to the origin of coordinates
in the plane. As it carries the same spin $\frac12$ particle the
lead component Hilbert space is $\HH_\mathrm{lead}=
L^2(\R_+,\C^2)$, and the whole state space of the system is the
consequently the orthogonal sum $\HH:= \HH_\mathrm{lead} \oplus
\HH_\mathrm{plane}$. The wave functions are thus of the form $\Psi
= \{ \psi_\mathrm{lead}, \psi_\mathrm{plane}\}^\mathrm{T}$ where
each of the components is a $2\times 1$ column.

Our aim is to find the dynamics of the particle on the described
configuration space. The idea of the construction is the same as
in \cite{ES87}, and since it was used many times --- let us just
recall \cite{BEG03, BG03, BGMP02, ES97} as a sample of this work
--- we can just recall briefly the scheme and describe how it
applies in the present situation. We start from the decoupled
operator $H^0:=H_\mathrm{lead} \oplus H_J$ where the first
component is the Laplacian on the halfline $H_\mathrm{lead}
\psi_\mathrm{lead} = -\psi''_\mathrm{lead}$ with Neumann boundary
condition at the endpoint, and $H_\rJ$ is the Hamiltonian with the
spin-orbit interaction discussed in the previous section. We
restrict $H^0$ to functions which vanish in the vicinity of the
junction, obtaining thus a symmetric operator of deficiency
indices $(4,4)$, and after that we seek admissible Hamiltonians
among its self-adjoint extensions.

The problem differs from those mentioned above only by the
presence the spin degree of freedom and the best way to
characterize the extensions is again through boundary conditions.
To this aim we need the boundary values. Those on the halfline are
the columns $\psi_\mathrm{lead}(0+)$ and $\psi'_\mathrm{lead}
(0+)$. In the plane the functions from the domain of the
restriction have a logarithmic singularity at the origin and the
generalized boundary values $L_j(\psi_\mathrm{plane}),\,
\mbox{j=0,1,}\,$ appear as coefficients in the expansion
 \[ \label{gen_bv}
\psi_\mathrm{plane}(\bx) = -\frac{1}{2\pi}\,
L_0(\psi_\mathrm{plane})\,\ln |\bx| + L_1(\psi_\mathrm{plane}) +
o(|\bx|)\,.
 \]
Using this notation we can write the sought boundary conditions as
 \[ \label{bc}
\begin{array}{rcl}
\psi'_\mathrm{lead} (0+) &=& A\psi_\mathrm{lead} (0+) +
C^*L_0(\psi_\mathrm{plane})\,, \\ [.3em] L_1(\psi_\mathrm{plane})
&=& C\psi_\mathrm{lead} (0+) + DL_0(\psi_\mathrm{plane})\,,
\end{array}
 \]
where $A,C,D$ are $2\times 2$ matrices, the first and the third of
them Hermitian, so the matrix $\AA:= {A\;C^* \choose C\;D}$
depends of sixteen real parameters as expected. It is
straightforward to check that the corresponding boundary form
vanishes under the condition (\ref{bc}), and therefore each fixed
$\AA$ gives rise to a self-adjoint extension $H_\AA$ of the
restricted operator. Notice that the analogous boundary conditions
apply also to the magnetic case mentioned in the previous section
due to the same character of the singularity.

Let us further mention that the above boundary conditions are
generic but do not describe all the extensions leaving out cases
when the matrix $\AA$ is singular; this flaw can be corrected in
the standard way -- see, e.g., \cite{AP05} -- if one replaces
(\ref{bc}) by the symmetrized form of the relation,
 \[ \label{bc2}
\AA {\psi_\mathrm{lead}(0+) \choose L_0(\psi_\mathrm{plane})} +
\BB {\psi'_\mathrm{lead}(0+) \choose L_1(\psi_\mathrm{plane})}
=0\,,
 \]
where $\AA,\BB$ are matrices such that $(\AA|\BB)$ has rank four
and $\AA\BB^*$ is Hermitean. We will restrict ourselves, however,
to the case $\BB=-I$ in the following; the same is true for the
alternative form of the b.c. mentioned below.

The choice of the parameter matrix depends on the way in which the
lead and the plain are connected. In particular, diagonal $A,C,D$
correspond to the situation when the contact does not couple the
spin states, and in addition, the matrices are scalar if the
coupling is spin-independent. Moreover, the lead and the plane are
decoupled if $\AA$ is block-diagonal, i.e. $C=0$. A na\"{\i}ve
interpretation is that $C$ is responsible for the coupling while
$A$ and $D$ are point perturbations at the contact ``from the two
sides''. The reality is more complicated, though. If the halfline
models a thin fibre of radius $\rho$ coupled to the plane, then in
the spin-indepedent case the natural choice seem to be
 \[ \label{nat}
A=\frac{1}{2\rho}\, \sigma_0\,, \quad
C=\frac{1}{\sqrt{2\pi\rho}}\, \sigma_0\,, \quad D= -\sigma_0\,
\ln\rho
 \]
in analogy with the discussion of then spinless case performed in
\cite{ES97}.

\setcounter{equation}{0}
\section{The Green function}

Having defined the class of admissible Hamiltonians we can proceed
to construction of their resolvents. One can use the standard
procedure based Krein's formula -- see, e.g., \cite{AGHH} or
\cite{BGMP02}. The starting point is Green function of the
decoupled system which has a block-diagonal form,
 \[ \label{freeG}
G^0(x,x';\bx,\bx';z)=\left(\begin{array}{cc}
G_\mathrm{lead}(x,x';z) & 0 \\
0 & G_{\rJ}(\bx,\bx';z) \end{array} \right)\,,
 \]
where $G_{\rJ}(\bx,\bx';z)$ is given by (\ref{SOGreen}) and
 $$
 G_\mathrm{lead}(x,x';z)= \frac{\rmi}{\sqrt{z}}\, \cos
 \sqrt{z}x_<\: \re^{-\rmi\sqrt{z}x_>}\: \sigma_0
 $$
with the conventional notation, $x_<:=\min\{x,x'\},\:
x_>:=\max\{x,x'\}$, since we assumed Neumann boundary condition.
The Krein function $Q(z)$, which is an analytic $4\times4$-matrix
valued function of the spectral parameter $z$, is defined through
diagonal values of the kernel, with the above described
renormalization in the plane component, specifically
 \[ \label{kreinf}
 Q(z):= \left(\begin{array}{cc}
 \frac{\rmi}{\sqrt{z}}\: \sigma_0 & 0 \\
 0 & G^{\mathrm{ren}}_\rJ(z) \end{array} \right)\,.
 \]
To express the full Green function it is useful to cast the
conditions of the previous section into an alternative form by
changing the basis in the boundary value space: instead of the
vectors appearing in (\ref{bc2}) we consider
 $$
 \tilde\Gamma_1\psi:= {-\psi'_\mathrm{lead}(0+) \choose
 L_0(\psi_\mathrm{plane})}\,, \quad \tilde\Gamma_2\psi:=
 {\psi_\mathrm{lead}(0+) \choose L_1(\psi_\mathrm{plane})}\,.
 $$
It is easy to see is that they satisfy $\tilde\AA\tilde\Gamma_1
\psi + \tilde\BB\tilde\Gamma_2 \psi=0$ with $\tilde\BB=-I$ and
 \[ \label{tildeA}
 \tilde\AA := \left(\begin{array}{cc}
 -A^{-1} & -A^{-1}C^* \\ -CA^{-1} & D-CA^{-1}C^*
 \end{array} \right)\,.
 \]
It is obvious that $\tilde\AA=-\tilde\AA\tilde\BB^*$ is Hermitean;
we have to suppose, of course, that the matrix $A$ is regular, or
roughly speaking, that $H_\AA$ has no Neumann component on the
halfline (notice that this is true, e.g., for (\ref{nat})). The
advantage of these boundary conditions is that our comparison
operator $H^0$ is characterized by $\tilde\Gamma_1 \psi=0$, i.e.
\mbox{$\tilde\AA^0=I$}, $\tilde\BB^0=0$. In such a case we can use
the result of \cite{AP05} (which in our case boils down to the
usual Krein's formula) by which the resolvent kernel of $H_\AA$ is
given by
 \begin{eqnarray} \label{krein}
 \lefteqn{G_\AA(x,x';\bx,\bx';z)= G^0(x,x';\bx,\bx';z)} \\ [.3em]
 && - G^0(x,0;\bx,\mathbf{0};z)\, \big[Q(z)-\tilde\AA\big]^{-1}
 G^0(0,x';\mathbf{0},\bx';z) \nonumber
 \end{eqnarray}
differing from the free one by the second term on the right-hand
side which is a rank sixteen operator. Notice that even if the
coupling is spin-independent, $\AA= {a\;\bar c \choose c\;d}
\otimes \sigma_0$ and similarly for $\tilde\AA$, the Green
function does not decompose because spin states are coupled by the
spin-orbit interaction in the plane.

\setcounter{equation}{0}
\section{Properties of $H_\AA$}

We will concentrate on the case when there is a nontrivial
coupling between the two parts of the configuration manifold, i.e.
$\AA$ is no block-diagonal. In the opposite case we have two
separate problems; the halfline one is trivial while spin-orbit
Hamiltonians with point interactions deserve an investigation --
we believe that the reader can find a study on this topic in the
contribution of K.~Pankrashkin to this issue. To keep things
simple, we suppose that the coupling is spin-independent, $\AA=
{a\;\bar c \choose c\;d} \otimes \sigma_0$ with $c\ne 0$, so
 \[ \label{krein_den}
 Q(z)= \left(\begin{array}{cc}
 \frac{\rmi}{\sqrt{z}}-\tilde a & -\tilde{\bar c} \\
 -\tilde c & G^{\mathrm{ren}}_\rJ(z)-\tilde d \end{array} \right)\,
 \otimes \sigma_0\,.
 \]

Let us first remark that the junction can bind. For instance, to
any number in $(-\vk_\rJ^2,0)$ one can find $H_\AA$ for which it
is an eigenvalue. Indeed, writing the negative energy as
$-\kappa^2$ we see that (\ref{krein_den}) is singular if the
relation $(\kappa^{-1}-\tilde a)(G^{\mathrm{ren}}_\rJ(-\kappa^2)
-\tilde d) =|\tilde c|^2$ is valid, or in the original parameters
 \[ \label{spect}
 (\kappa-a)\big(G^{\mathrm{ren}}_\rJ(-\kappa^2)-d \big) =|c|^2\,.
 \]
By (\ref{Gren2}) $\,G^{\mathrm{ren}}_\rJ(-\kappa^2)$ is
real-valued for $\kappa^2<\vk_\rJ^2$, then it is easy to pick the
parameters $a,d$ in such a way that (\ref{spect}) is satisfied.

The most interesting aspect of the problem, of course, is the
transport through the junction. A straightforward way to treat it
is to use the formula (\ref{krein}). Any vector of $\HH$ can be
written as $(H^0-z)^{-1}\psi^0$ for $\psi^0\in D(H^0)$ and $\im
z\ne 0$, hence applying the formula to it we get
 \[ \label{psi}
 \psi = \psi^0 - \gamma_z [Q(z)-\AA]^{-1} \gamma_{\bar z}^*
 (H^0-z)^{-1}\psi^0\,,
 \]
where $\gamma_z:\: \C^4\to \HH$ is the trace operator given by the
kernel $G^0(x,0;\bx,\mathbf{0};z)$ and $\gamma_z^*$ is its
adjoint. Notice that $\Gamma(\bar z)^*(H^0-z)^{-1}\psi^0$ is just
the vector of the values of $\psi^0$ at the connection point and
$Q(z)-\AA$ is position-independent, so the second term at the
right-hand side is easy to compute. Now we employ the usual trick
letting $z$ to approach a real value $k^2$. The resulting function
ceases to be $L^2$, of course, but it still satisfies locally the
boundary conditions in the junction and it can yield a generalized
eigenfunction associated with the scattering which we are looking
for.

In particular, we can choose the vector $\psi^0$ with the
``upper'' component only, $\psi^0_\mathrm{plane}=0$ and
$\psi^0_\mathrm{lead}=\cos kx\:$ (notice that not every
combination of $\re^{\pm\rmi kx}$ will do since
$\psi^0_\mathrm{lead}$ has to satisfy Neumann boundary condition
at the origin). It is a straightforward exercise to invert the
matrix (\ref{krein_den}) and to compute $\psi\,$; it yields the
reflection amplitude of a particle travelling over the halfline
towards the junction with a momentum $k$ in the form
 $$
 \RR(k)= \frac{\left(-\frac{\rmi}{k}-\tilde a\right)
 (G^{\mathrm{ren}}_\rJ(k^2)-\tilde d)-|\tilde c|^2}
 {\left(\frac{\rmi}{k}-\tilde a\right)
 (G^{\mathrm{ren}}_\rJ(k^2)-\tilde d)-|\tilde c|^2}\,,
 $$
naturally independent of the particle spin state. It is
straightforward to recompute it in terms of the original
parameters from (\ref{bc}); we get
 \[ \label{refl}
 \RR(k)= -\,\frac{(a+\rmi k)
 (G^{\mathrm{ren}}_\rJ(k^2)-d)+|c|^2}
 {(a-\rmi k) (G^{\mathrm{ren}}_\rJ(k^2)-d)+|c|^2}\,.
 \]
Since $G^{\mathrm{ren}}_\rJ(k^2)$ is generally complex
$|\RR(k)|^2$ is not equal to one for $|c|\ne 0$ which is natural
because the coupling allows the particle to pass from the lead to
the plane. In particular, in the absence of the spin-orbit
coupling when the last term at the right-hand side of
(\ref{Gren2}) is missing, (\ref{refl}) reduces to the reflection
amplitude derived in \cite{ES87} for spinless case (up to the sign
of $k$ which is due to the opposite halfline orientation used
there).

In the magnetic case one can proceed in the same way replacing
$G^{\mathrm{ren}}_\rJ(k^2)$ by the renormalized magnetic Green's
function of \cite{BGP07}. There is a substantial difference,
though. Now the Green function is real-valued, and consequently,
the scattering on the halfline is unitary, $|\RR(k)|^2=1$. The
scattering in such a case will naturally exhibit resonances due to
the discrete spectrum of the spin-orbit Hamiltonian in the plane
which are worth of investigation, however, this is a subject for
another paper.

\subsection*{Acknowledgments}

The research was supported in part by the Czech Academy of
Sciences and Ministry of Education, Youth and Sports within the
projects A100480501 and LC06002. The first author appreciates the
hospitality in the Isaac Newton Institute in Cambridge where a
part of this work was done.


\begin{thebibliography}{99}

 \bibitem[AGHH]{AGHH}
 S.~Albeverio, F.~Gesztesy, R.~H\o egh-Krohn, H.~Holden:
 \emph{Solvable Models in Quantum Mechanics}, 2nd edition, with an
 appendix by P.~Exner, AMS Chelsea 2005.
 \vspace{-1.8ex}
 \bibitem[AP05]{AP05}
 S.~Albeverio, K.~Pankrashkin: A remark on Krein's resolvent formula
 and boundary conditions, \emph{J.Phys.A: Math. Gen.}
 \textbf{38} (2005), 4859--4864.
 \vspace{-1.8ex}
 \bibitem[BEG03]{BEG03}
 J.~Br\"uning, P.~Exner, V.A.~Geyler: Large gaps in point-coupled
 periodic system of manifolds, \emph{J. Phys. A: Math. Gen.}
 \textbf{36} (2003), 4875--4890.
 \vspace{-1.8ex}
 \bibitem[BG03]{BG03}
 J.~Br\"uning, V.A.~Geyler: Scattering on compact manifolds with
 infinitely thin horns, \emph{J. Math. Phys.} \textbf{44} (2003), 371--405.
 \vspace{-1.8ex}
 \bibitem[BGMP02]{BGMP02}
 J.~Br\"uning, V.A.~Geyler, V.A.~Margulis, M.A.~Pyataev: Ballistic
 conductance of a quantum sphere, \emph{J. Phys. A: Math. Gen.}
 \textbf{35} (2002), 4239--4247.
 \vspace{-1.8ex}
 \bibitem[BGP07]{BGP07}
 J.~Br\"uning, V.~Geyler, K.~Pankrashkin: Explicit Green functions
 for spin-orbit Hamiltonians, \texttt{arXiv:0704.2877v1 [math-ph]}
 \vspace{-1.8ex}
 \bibitem[E\v{S}87]{ES87}
 P.~Exner, P.~\v Seba: Quantum motion on a halfline connected to a
 plane, \emph{J. Math. Phys.} \textbf{28} (1987), 386--391; erratum
 p.~2254.
 \vspace{-1.8ex}
 \bibitem[E\v{S}97]{ES97}
 P.~Exner, P.~\v Seba: Resonance statistics in a microwave
cavity with a thin antenna, \emph{Phys. Lett.} \textbf{A228}
(1997), 146--150.
 \vspace{-1.8ex}





\end{thebibliography}
\end{document}